# Towards broadband dynamic structuring of a complex plasmonic field

Shibiao Wei,[1,2,3] Guangyuan Si,[3] Michael Malek,[2] Stuart K. Earl,[2,3] Luping Du,[1] Shan Shan Kou,[2] Xiaocong Yuan[1,*] and Jiao Lin[1,3,*]

[1]*Key Laboratory of Optoelectronic Devices and Systems of Ministry of Education and Guangdong Province, Nanophotonics Research Center, College of Optoelectronic Engineering, Shenzhen University, Shenzhen 518060, China*

[2]*Department of Chemistry and Physics, La Trobe Institute for Molecular Science (LIMS), La Trobe University, Victoria 3086, Australia.*

[3]*School of Engineering, RMIT University, Melbourne, Victoria 3001, Australia.*

[*]Correspondence to:  xcyuan@szu.edu.cn or jiao.lin@osamember.org

**The ability to tailor a coherent surface plasmon polariton (SPP) field is an important step towards a number of new opportunities for a broad range of nanophotonic applications such as sensing [1,2], nano-circuitry [3,4], optical data storage [5,6], super-resolution imaging [7,8], plasmonic tweezers [9,10] and in-plane communications [11]. Scanning a converging SPP spot or designing SPP profiles using an ensemble of spots have both been demonstrated previously [12-14]. SPPs, however, are normally excited by intense, coherent light sources—lasers. Hence, interference between adjacent spots is inevitable and will affect the overall SPP field distributions. Here we report a reconfigurable and wavelength-independent platform for generating a tailored 2D SPP field distribution by considering the coherent field as a whole rather than individual spots. The new perspective also enables us to reveal the inherent constraints implied in a 2D coherent field distribution. Our generic design methodology works not only for SPP waves but also for other two-dimensional wave systems like surface acoustic waves [15].**

To exist, in-plane SPP waves have to satisfy the Helmholtz equation, as do all electromagnetic waves, regardless of their dimensionality [16]. Due to the tight confinement at a metal-dielectric interface, the propagation of SPPs can be modeled in a 2D wave equation by 'losing' one of the dimensions as compared to electromagnetic waves propagating in free space. At a metal-dielectric interface in the x-y plane, the out-of-plane electric field component $E_{z,d}$ (in the dielectric half-space) is the most common component one used to characterize the 2D field distribution. For a monochromatic SPP, the field distribution of $E_{z,d}$ is governed by the 2D wave equation

$$\frac{\partial^2 E_{z,d}}{\partial x^2} + \frac{\partial^2 E_{z,d}}{\partial y^2} + k_{spp}^2 E_{z,d} = 0, \tag{1}$$

where $k_{spp} = k_0 \sqrt{\varepsilon_m \varepsilon_d /(\varepsilon_m + \varepsilon_d)}$ is the vector describing the SPP waves at the interface, $k_0$ is the wave vector in vacuum, and $\varepsilon_m, \varepsilon_d$ are the permittivities of the metal and dielectric, respectively. A number of SPP field distributions that satisfy the equation have been investigated as the 2D counterparts of free space solutions to the wave equation. These include in-plane self-accelerating beams [17,18], diffraction-free beams [19,20] and 2D plasmonic vortex beams [21]. Dynamically scanning an SPP focal spot (a point-like solution) has previously been demonstrated [12]. Intricate intensity patterns have also been generated by carefully arranging the in-plane converging spots of a number of SPPs [13,14]. These SPP spot arrays are possible to be generated because a converging spot is a solution to Equation (1), with its amplitude profile described by a zeroth-order Bessel function of the first kind [22]. However, the Bessel-function amplitude profile has many side-lobes, and therefore the entire SPP field must be assembled using discrete points with a reasonable spatial separation. In contrast, in this paper we demonstrate the full tailoring of complex SPP fields as a single inseparable entity. By considering an SPP field as a whole the complex fields, including amplitude, phase, and subsequently the in-plane energy flow of the entire 2D field distribution, can be manipulated. Furthermore, by using a wavelength-independent sub-wavelength structure to excite the SPP waves, we demonstrate the broadband nature of this approach. Experimentally, an electronically-addressed spatial light modulator (SLM) is used to imprint information onto the incident beam prior to the excitation of the designed SPP field. Hence, the resultant SPP fields become digitally controllable, which will facilitate many novel interdisciplinary applications that require a dynamically reconfigurable SPP field.

Our aim was to launch tailored monolithic SPP fields on a metal surface by projecting beams with a specifically-designed launching condition onto the ring coupler (depicted in Figure 1), a sub-wavelength annular ring-groove functions as an in-plane coupler to convert light propagating in free space to SPP waves. The experiment is depicted in Figure 1. Here, we use an iterative algorithm that involves a 2D in-plane Fourier transform (FT) operator inside the loop for retrieving the initial launching condition at the annular ring source. (For more details of algorithm see the Supplementary Information S1 and S2). To simplify the design procedure, the amplitude distribution of the initial launching condition at the ring coupler was set to be uniform during the iteration. The treatment is similar to the widely used Gerchberg–Saxton (GS) algorithm in phase retrieval [23]. The resultant phase distribution along the ring is subsequently transformed into a 2D radial pattern that was loaded onto the SLM. A free space, collimated laser beam was reflected from the SLM, imprinted with the designed phase distribution (also known as the initial launching condition) before it was projected on to the ring coupler to launch the desired SPP field. This was achieved by a 4-f system, as shown in Figure 1. As SPP waves can only be excited by transverse magnetic (TM) electromagnetic radiation, a radially polarized beam would, ideally, be used to launch the SPP waves from the ring. In the experiment, a circularly polarized beam was used instead, as the generation of a radially-polarized beam presented additional complexity. To cancel out the unwanted geometric phase generated by the coupling of circularly polarized light to the SPPs, a spiral phase was added to the phase distribution loaded on the SLM [24] (more details can be found in the Supplementary Information S3).

An aperture-type near-field scanning optical microscope (NSOM) (NT-MDT NTEGRA Solaris) with an aluminum coated fiber tip (tip aperture approximately 100 nm in diameter) was used in collection mode to measure the near-field intensity profile generated by the SPPs. Figure 2 shows experimental measurements which include a ring-shaped intensity pattern, elliptically-shaped intensity patterns with different eccentricities and directions, and a petal-shaped intensity pattern (experiment set up and more results can be found in the Supplementary Information S4 and S5). It is worth pointing out that the petal-shaped intensity pattern (Figure 2(d1) to (d4)) cannot be generated by simply overlapping two orthogonal elliptical intensity patterns because of the interference between them.

The dynamic control of a 2D phase distribution is shown in Figure 3. For example, it is possible to change the 'direction of phase gradient' from the 'forward' to the 'reverse' direction (the naming of these directions is defined arbitrarily) without altering the intensity distribution because they are two degeneracy solutions of the Equation (1). This would allow one to manipulate the in-plane energy flows in either the forward or reverse direction.

Often, structured SPPs are launched by complex nanostructured arrays [13,14,25]. The nanostructured units are placed over a specific interval designed for a certain wavelength. Normally, nanostructures are unable to generate SPPs in a broadband manner due to the wavelength-specific configuration. In contrast, the structure proposed here is a wavelength-independent ring-groove whose only function is to couple the incident light (which itself carries the whole information that defines the SPP field afterwards) from free-space into the in-plane SPPs. Figure 4 shows the numerical results of a variety of structured SPP fields using several incident wavelengths to demonstrate the broadband behavior of our approach.

The 2D field profile of a coherent monochromatic SPP consists of two parts: phase and amplitude. A question that may arise is whether we can control the in-plane phase and amplitude of an SPP field concurrently, the answer to which is dependent on whether the desired field distribution satisfies equation (1). When considering one of the parts (normally amplitude) by relaxing the distribution of the other part (e.g. phase), it is easier to find a resultant field that both satisfies the wave equation and approximates the target field distribution (e.g. amplitude distribution). Unfortunately, an arbitrary complex amplitude (both amplitude and phase) distribution of an in-plane electromagnetic wave does not necessarily satisfy equation (1). Take, for example, the plasmonic vortex beam field. The optical vortex (OV) field carried an optical angular momentum, which comprise concentric rings and, more importantly, have an associated azimuthal phase distribution [26]. In free space, the radii of the intensity profile for an OV beam can be varied [27]. In a 2D system such as an SPP, the size of a plasmonic vortex (PV) becomes unchangeable, as they are correlated to topological charge [28]. The eigenmodes of PVs on a gold film are $E_{l,z}(\varphi,r) \propto J_l(k_{spp}r)\exp(il\varphi)$, where $l$ is an integer representing the topological charge of the vortex. $J_l$ stands for the $l$ th-order Bessel function of the first kind[29]. The radius of the main intensity profile of PV field is described by the first peak of the Bessel function. It indicates an SPP wave circulates along this ring with a phase change of $l \times 2\pi$, therefore, intuitively the circumference of the ring is fixed to $l \times \lambda_{spp}$. The results obtained by our method, presented in Figure 5, also confirm that it is unable to find an appropriate output field whose radius does not match to correct value. For instance, the PV field, pictured in Figure 5(e)-(h), with a target ring-shape intensity profile intentionally larger than that of Figure 5(a)-(d)), could not satisfy Equation (1). Therefore the output SPP field resulted from our algorithm no longer resemble the target, in other words, the unphysical target SPP field could not be generated at a metal-dielectric interface. .

In summary, we have reported a new and versatile platform for dynamically controlling the in-plane SPP fields. We have demonstrated the modulation and generation of a number of intensity patterns by considering the coherent SPP field as a single entity. In addition, this technology could dynamically control the 'phase directions' of whole-SPP fields, which is the same as the direction of the (in-plane) energy flow.  The combination of a well-designed intensity profile and in-plane energy flow has the potential for dynamically controlling nanoparticles using plasmonic tweezers. Because the phase information required to launch the SPP field is carried by the incident laser beam without involving the design of wavelength-dependent nanostructures, this method is can be used for a range of wavelengths. This broadband ability offers potential applications in generating SPP fields with different colors, for high-speed in-plane communications, and for use in large-capacity data storage. Finally, the limits of two-dimensional modulation of SPP fields discussed here could act as a guide for the manipulation of other surface-confined waves, such as the modulation of surface acoustic waves.

**Method**

**Numerical simulation.** All of the numerical simulation work was performed with the commercial software FDTD Solutions from Lumerical Solutions, Inc. In the simulation, the dielectric constant of the gold film was from the built-in material database within the software, in which the raw data is taken from ref. 30, and the refractive index of glass was set to be 1.51; Perfectly matched layer (PML) boundaries were employed for the X, Y and Z directions. A global mesh size of 20 nm was applied in the calculations.

**Sample preparation.** A 100 nm gold film was deposited onto a quartz substrate by electron-beam evaporation with a 5 nm germanium adhesion layer. The ring-groove structures were fabricated using focused ion beam milling (FEI Helios Nanolab600 Dual Beam FIB-SEM system). During FIB milling a 28 pA beam current was selected with an accelerating voltage of 30 kV.

**Phase control.** The algorithm discussed earlier provides only a single output for the intensity target, which has two degenerate phase distributions that differ by their 'phase direction'. Both solutions have the same probability of being found by the iterative process of the algorithm, and the initial conditions determine which one is identified. It is possible to get the other solution for the target field by changing the initial conditions of the algorithm. Our method of retrieving the second solution for the target field is by reversing the phase distribution of the first solution and using this as the initial conditions of the algorithm.

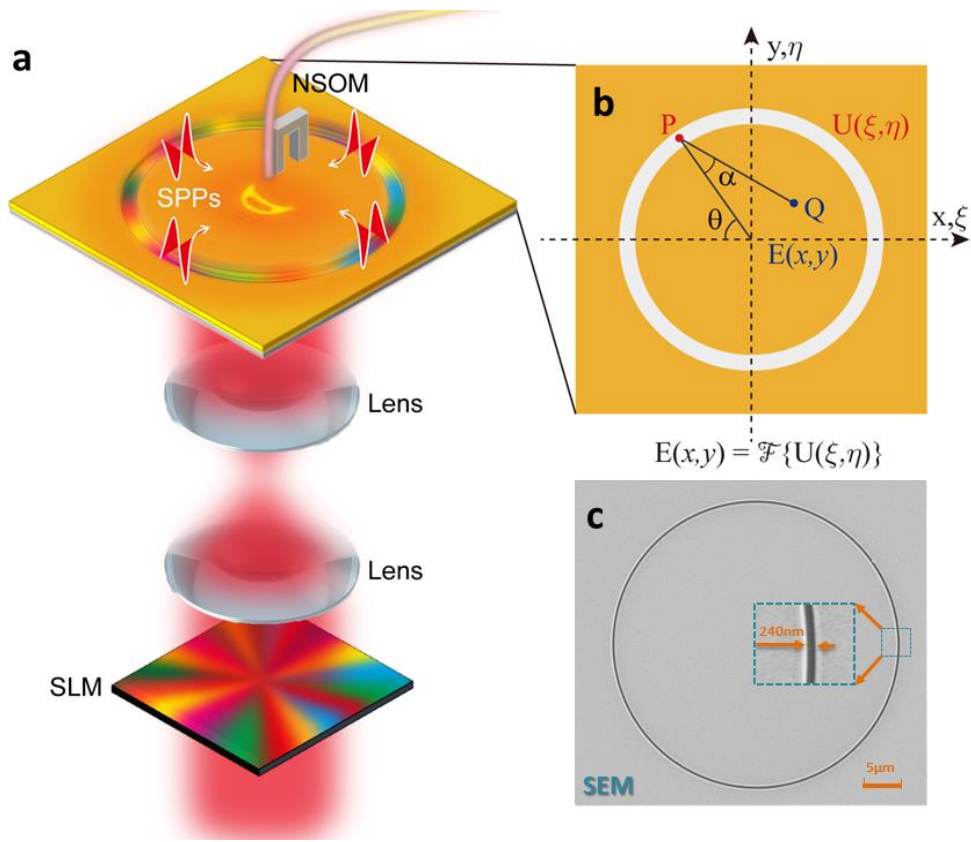

**Figure 1| Algorithm and schematic diagram.** (a) Schematic diagram of the SPP field launching method. A 4-f system was constructed using an achromatic doublets lens (f = 500 mm), and an objective lens (OLYMPUS, LUCPlan FLN, 20×, NA=0.45) was used to project the phase distribution (loaded on the SLM) to the structure milled on the gold surface. A near-field scanning optical microscope working in collection mode was used to measure the SPP fields on the gold surface. (b) In the two-dimensional plane, a 2D Fourier relationship can be found for the complex amplitude along a ring $U(\zeta,\xi)$ of a converging surface wave and the in-plane component of the SPP field $E_z(x, y)$ in the vicinity of the geometrical focus. After the iterative loops, a phase distribution of the initial launching condition for the ring source could be obtained for loading onto the SLM (details could be found in Supplementary Information). (c) SEM micrograph of the annular ring-groove nanostructure. The diameter of the ring is $40\mu m$, while the width of the groove is 240nm. The scale bar is $5\mu m$.

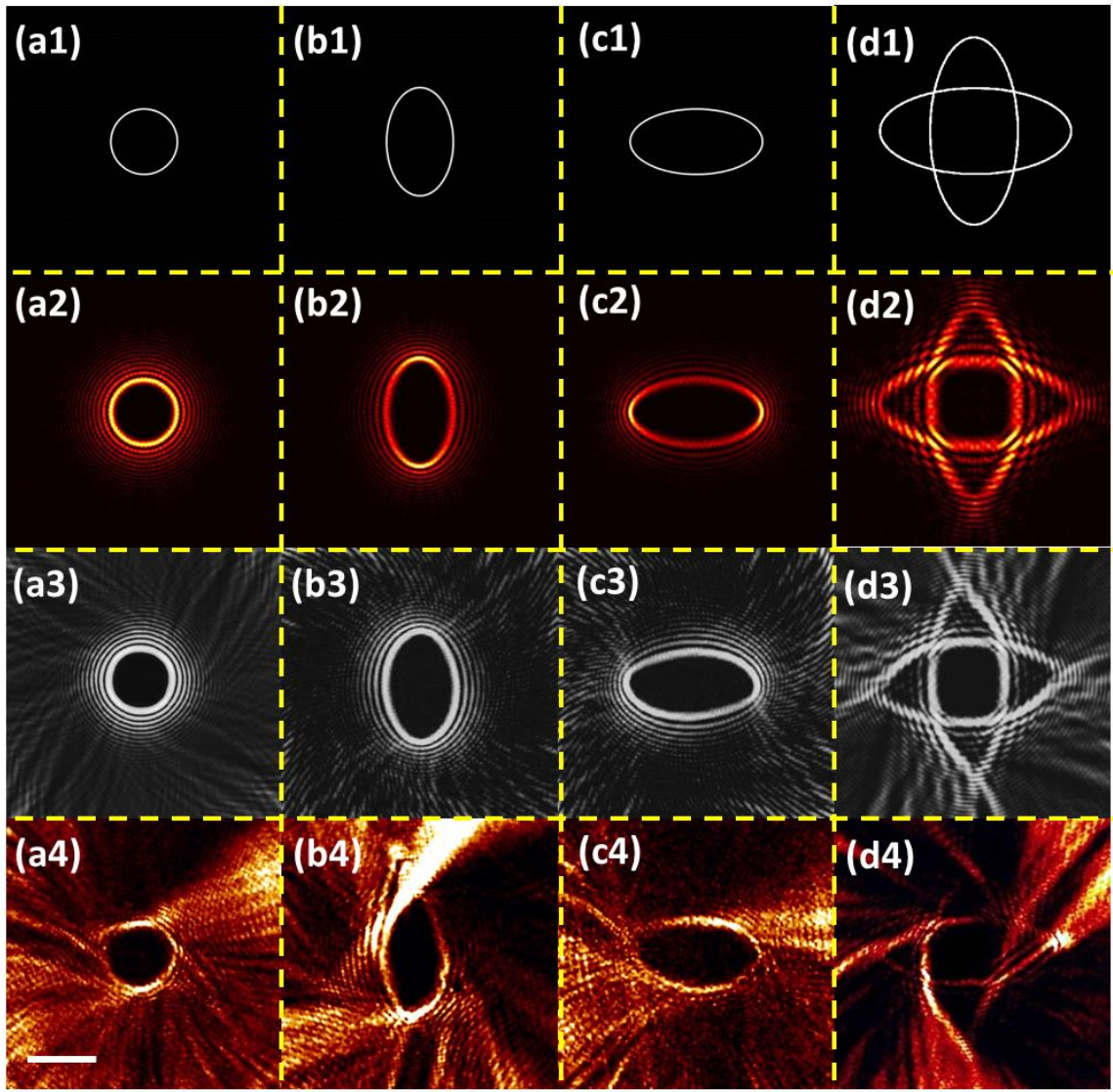

**Figure 2| Results of dynamic and arbitrary control of SPP fields.** Images (a1) to (d1) the target fields of what we want to generate on the gold surface. Images (a2) to (d2) are the Fourier transform results of the initial ring phase distributions which were obtained using the algorithm. Images (a3) to (d3) are the FDTD simulation results using the initial ring phase distributions to excite SPP waves. Images (a4) to (d4) are NSOM measurements. The scale bar is $5 \mu m$.

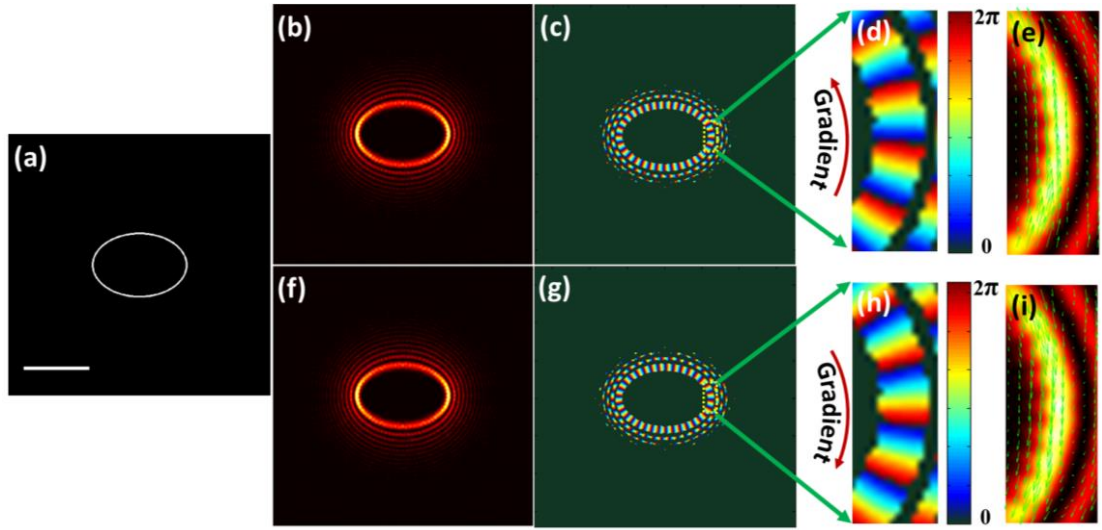

**Figure 3| In-plane phase control.** (a) The ellipse intensity target is what we want to generate. (b), (c) and (f), (g) are intensity and phase distributions of the SPP field obtained using the launch conditions calculated using our algorithm. One can see the intensity profiles presented in (b) and (f) are the same. (d) and (h) Close-up image of the yellow dashed square shown in (c) and (g), showing that the 'phase directions' are reversed, indicatung that the propagation direction of the SPP waves are reversed. Here, we have set the phase distribution to zero in the low-intensity areas. (e) and (i) Poynting vectors of the two field, indicating reversed energy flow. The scale bar is $5\mu m$.

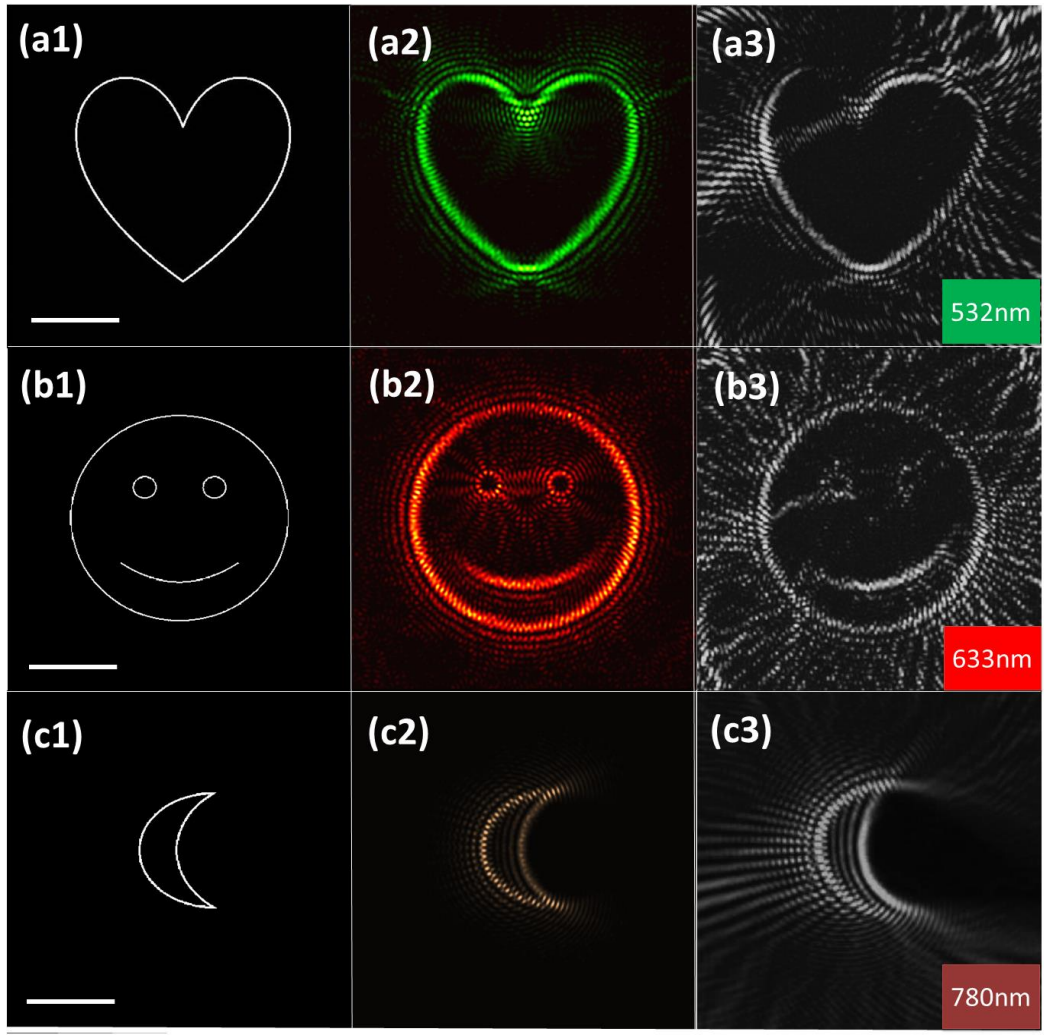

**Figure 4| Schematic and results using broadband excitation of SPP fields.** Images (a1) to (a3) are intensity targets showing what we want to generate, Fourier transform results of the initial ring phase distributions which were obtained by the algorithm, and FDTD simulation results using the initial ring phase distributions to excite SPP waves, respectively, when the wavelength of the incident beam is 532nm. Images from (c1) to (c3), and images from (d1) to (d3) are identical results for different SPP distributions when the wavelength of the incident beams are 633nm and 780nm, respectively. The scale bar is $5\mu m$.

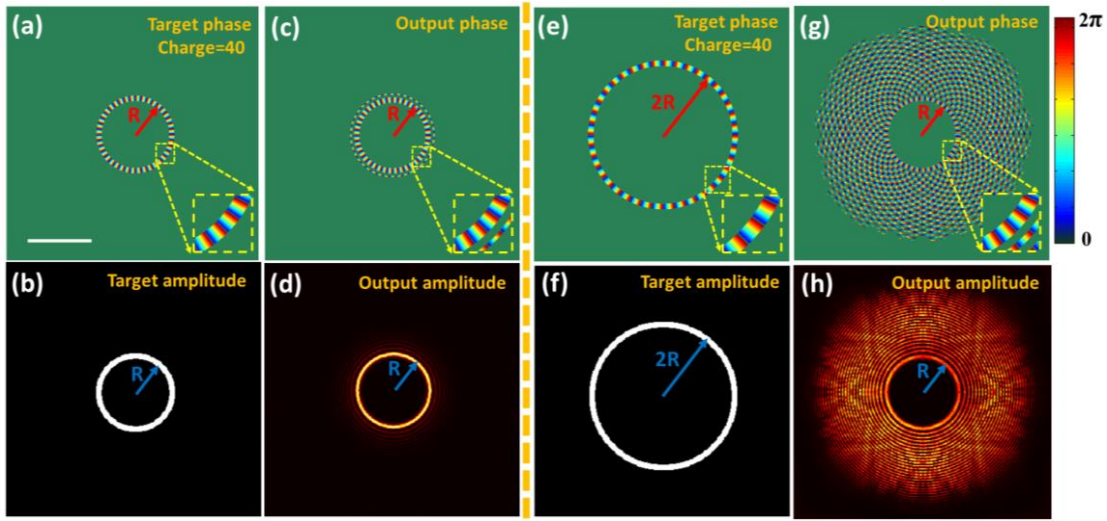

**Figure 5| In-plane SPP spiral phase field generation.** (a)Spiral phase distribution of the target PV with a topological charge value of 40. (b) The target intensity distribution of the PV field is set to the same size as the phase distribution. (c) and (d) The intensity distribution and phase distribution of Fourier transform results of the initial ring phase distributions, which were obtained using our algorithm. One can see a PV field with a topological charge of 40 can be generated. (e) Spiral phase distribution of the target PV field with the same topological charge of 40. (f) The radius of the ring of the intensity distribution is double the size of (a). (g) and (h) Fourier transform results of the initial ring phase distributions, which were obtained using our algorithm, showing this target PV field could not be generated on the metal surface for the increased intensity ring shown in (f). The energy still flows to the ring position with the radius of R generating a PV field whose charge equal to 40. But the amplitude in (f) is much smaller than that in (d). The scale bar is $5\mu m$.

# Supplementary information

## S1: In-plane Fourier transform for a converging SPP waves

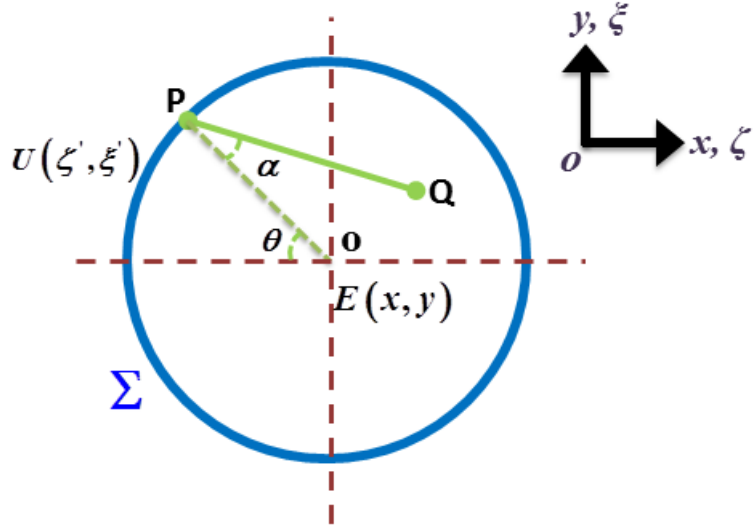

**Figure S1** Derivation of the in-plane Fourier transform of surface waves. The ring of $\mathsf{S}$ presents the convergent wave front.

For monochromatic propagating SPPs at a source-free metal/dielectric interface (z=0) in the xy plane, the in-plane SPP wave vector could be described by $k_{spp} = \sqrt{k_x^2 + k_y^2} = k_0\sqrt{\varepsilon_m \varepsilon_d/(\varepsilon_m + \varepsilon_d)}$, where $k_0$ is the wave vector in vacuum, $\varepsilon_m$ is the permittivity of the metal, $\varepsilon_d$ is the permittivity of the dielectric. The longitudinal component of electric field $E_z$ should obey:

$$\frac{\partial^2 E_z}{\partial x^2} + \frac{\partial^2 E_z}{\partial y^2} + k_{spp}^2 E_z = 0. \tag{S1}$$

We apply the Weber's analogue integral theorem of Helmholtz-Kirchhoff[31], the disturbance at $U$ (Fig. S1) for a convergent SPP wave can be given by

$$E_z(x,y) = -\frac{i}{4}\int_\Sigma \left\{ U\frac{\partial H_0^{(1)}(k_{spp}d)}{\partial n} - H_0^{(1)}(k_{spp}d)\frac{\partial U}{\partial n} \right\} ds, \tag{S2}$$

where $U$ is the field distribution on the closed curve consisting of $\mathsf{S}$; $d$ is the distance between $P$ and $Q$; $\frac{\partial}{\partial n}$ means the differentiation along the outward normal to the closed curve; $H_m^{(1)}(x)$ is the $m^{\text{th}}$ order Hankel function of the first kind. By using $ds = Rd\theta$, $\frac{d}{dx}H_0^{(1)}(x) = -H_1^{(1)}(x)$ and $\frac{d}{dx}H_0^{(2)}(x) = -H_1^{(2)}(x)$ where R is the radius of the ring, $H_m^{(2)}(x)$ is the *m*th order Hankel function of the second kind [32], we obtain:

$$E_z(x,y) = i\frac{k_{spp}R}{4}\int_\Sigma \left\{H_0^{(2)}(k_{spp}R)H_1^{(1)}(k_{spp}d)\cos\alpha - H_1^{(2)}(k_{spp}R)H_0^{(1)}(k_{spp}d)\right\}U(\theta)d\theta. \tag{S3}$$

For the field distribution near the focus, $\alpha \approx 0$. If $R \gg \lambda_{spp}$ and $d \gg \lambda_{spp}$, Eq. (S3) can be simplified as:

$$E_z(x,y) \approx \frac{e^{-ik_{spp}R}\sqrt{R}}{\pi}\int_\Sigma U(\theta)\frac{e^{ik_{spp}d}}{\sqrt{d}}d\theta \tag{S4}$$

which can be regarded as a 2D integral:

$$E_z(x,y) = \int_{-\pi}^{\pi}\int_0^\infty U(\theta)\delta(r-R)\frac{e^{ik_{spp}d}}{\sqrt{d}}rdrd\theta. \tag{S5}$$

$\delta(\cdot)$ is the Dirac delta function. By neglecting the imaginary part of $k_{spp}$ since it is much smaller than the real part and using the following approximation:

$$d = \sqrt{(\xi-x)^2 + (\zeta-y)^2} = \sqrt{R^2 + x^2 + y^2 - 2(\xi x + \zeta y)} \approx R\sqrt{1 - \frac{2}{R^2}(\xi x + \zeta y)}$$

$$\approx R\left[1 - \frac{1}{R^2}(\xi x + \zeta y)\right] = R - \frac{\xi x + \zeta y}{R} \tag{S6}$$

we have:

$$E_z(x,y) = \int_{-\pi}^{\pi}\int_0^\infty U\delta(r-R)\exp\left[-i\frac{k_{spp}}{R}(\xi x + \zeta y)\right]rdrd\theta$$

$$\approx \int_{-\infty}^{\infty}\int_{-\infty}^{\infty}U'(\xi',\zeta')\exp\left[-i2\pi(\xi' x + \zeta' y)\right]d\xi'd\zeta', \tag{S7}$$

where $U'(\xi',\zeta') = U(\theta)\delta(r-R)$; $\xi' = \frac{\text{Re}(k_{spp})}{2\pi R}\xi = \frac{1}{R\lambda_{spp}}\xi$, $\zeta' = \frac{\text{Re}(k_{spp})}{2\pi R}\zeta = \frac{1}{R\lambda_{spp}}\zeta$.

Neglecting the constant factor gives:

$$E_z(x,y) = \mathcal{F}_2\{U'(\xi',\zeta')\} \tag{S8}$$

## S2: Algorithm for obtaining the initial-launching-distribution of SPP fields.

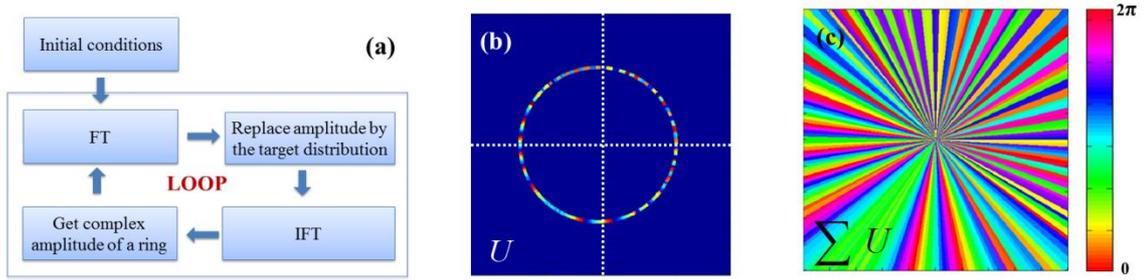

Figure S2 Algorithm processing flows to obtain the initial-launch-distribution of SPP fields. (a) The iterative processing flows of the algorithm. (b) The ring-phase pattern $U$ of the initial-launching-distribution. (c) Finalized phase distribution $\sum U$ extended along the radial direction of ring-phase pattern $U$.

In this section, we show how to obtain the initial-launch-distribution which is loaded onto the SLM to excite specific SPP fields. Fig. S2 (a) shows the process flow to obtain a ring-shaped distribution of the initial-launch-distribution for an SPP field. The processing steps are

0. At the beginning, we define an initial condition for the starting loop of the algorithm. Here, a convergent wave front with ring-shape distribution is used, which could generate a focused spot at the center of the structure.
1. A forward Fourier transform is applied to the ring-shaped distribution. It is the processing of waves which propagate from the 'structure domain' to the 'target domain' for reconstructing the in-plane field on the surface of the gold film.
2. In the 'target domain', the amplitude term of the reconstructed field is replaced by the amplitude distribution of the target field with the phase term unchanged. A new complex amplitude distribution is generated.
3. An inversed Fourier transform is applied to the new complex amplitude distribution to produce a complex field distribution from the 'target domain' to the 'structure domain'.
4. In the 'structure domain', extract the complex amplitude or phase-only field located on the ring while zeroing the other regions to obtain a ring-shaped distribution.
5. The process is repeated from step 1.

The resulting reconstructed field (step 1) is compared with the target one. By using the correlation between both images as a criterion, a decision is taken to finish the process or continue iterating. To simplify the projecting process from the SLM to the sample plane, the phase distribution, labeled by field $U$ as shown in Fig. S2 (b), is extended to a spoke-like radial phase distribution, labeled by field $\sum U$ as shown in Fig. S2 (c).

## S3: Compensating spiral phase carried by the circularly polarized beam.

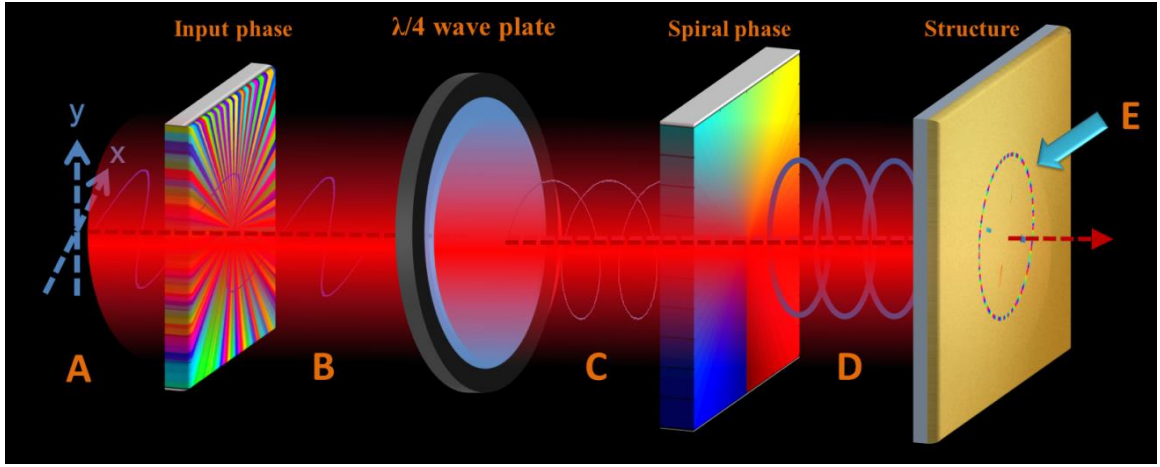

Figure S3 Schematic diagram of compensating spiral phase by an inversed spiral phase distribution.

SPP waves can be excited by transverse-magnetic (TM) polarized light at a dielectric/metal interface. A radially polarized beam (RPB) is an ideal information carrier for a ring-groove structure because the local electric field of an RPB is linearly polarized along the radial direction, and hence, fully TM polarized light an axially-symmetric structure. However, the processing of generating an RPB is complex and involves more optical components, which would significantly increase the complexity of the system. In addition, the polarized singularity at the center of the beam needs to be carefully aligned with the center of the both hologram and annular ring-groove. This also increases the complexity involved in adjusting and aligning the optical path. As a result, a circularly polarized beam (CPB) was selected for the experiment instead. In a cylindrical coordinate system, a left-hand circular (LHC) polarized beam with a planar wavefront can be expressed as

$$E_{LHC} = \frac{(e_x + ie_y)}{\sqrt{2}} = \frac{\left[(\cos\varphi e_r - \sin\varphi e_\varphi) + i(\sin\varphi e_r + \cos\varphi e_\varphi)\right]}{\sqrt{2}} = \frac{e^{i\varphi}(e_r + ie_\varphi)}{\sqrt{2}}. \quad (s9)$$

For simplicity, the CPB is assumed to have unit amplitude. The CPB can be decomposed into radially polarized and azimuthally polarized components with a spiral phase wave front. The azimuthally-polarized component can be ignored because it is a transverse-electric (TE) mode with respect to the ring-groove. The spiral phase carried by the radially-polarized component can be canceled out by the inclusion of an additional spiral phase, as shown in Figure S3. The polarization states corresponding to the positions denoted by capital letters A, B, C, and D in Figure S3 are

A: $\vec{e}_x$

B: $\sum U(r,\varphi)\vec{e}_x$

C: $\sum U(r,\varphi)e^{i\varphi}(\vec{e}_r + i\vec{e}_\varphi)$

D: $\sum U(r,\varphi)(\vec{e}_r + i\vec{e}_\varphi)$

E: $U(r,\varphi)\vec{e}_r$.

**S4:** Experimental setup

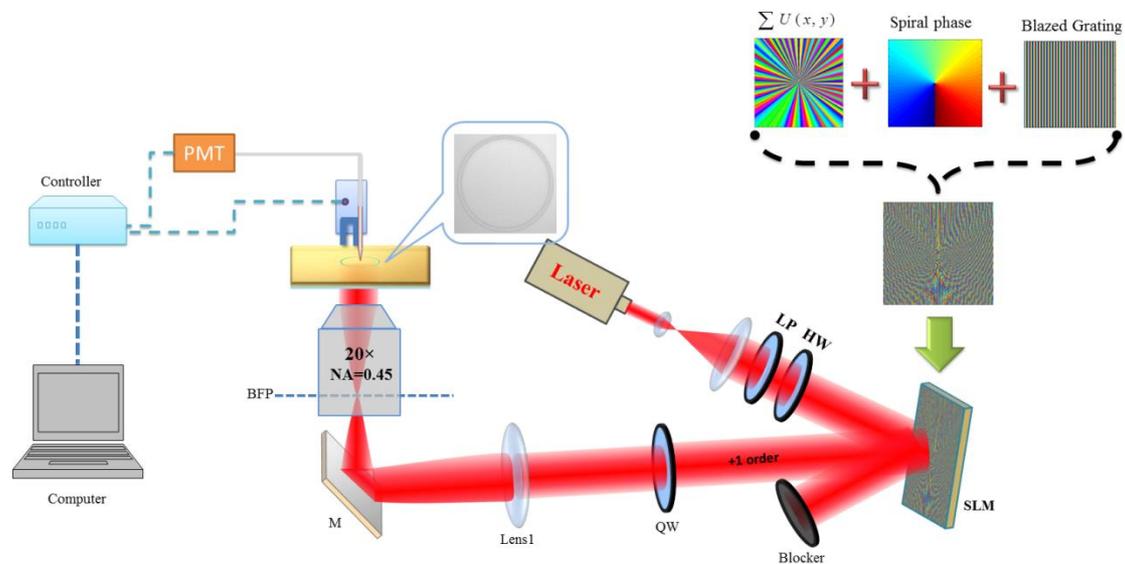

Figure S4 (LP: linear polarizer; HW: *λ/2* waveplate; QW: *λ/4* waveplate; SLM: spatial light modulator; M: mirror; BFP: back focal plane; PMT: photo-multiplier tube). Schematic diagram of the experimental setup.

A Helium-Neon laser (Melles Griot, 25-LHP-828) with a wavelength of $\lambda=632.8$nm was used. The linearly polarized laser beam was expanded and collimated by a telescope system and subsequently passed through a linear polarizer and a half waveplate to produce the desired polarization orientation and power. The linearly polarized laser beam illuminated the spatial light modulator (SLM, Holoeye PLUTO), which was used to imprint the designated phase pattern onto the reflected beam. Because the fill factor of the SLM was 87%, a blazed grating was added to eliminate the disturbance of the DC component of the incident beam which was not modulated by the liquid crystal molecules of the SLM. To achieve this, the phase patterns loaded on the SLM were generated by overlapping three components: an input phase $\sum U$, a spiral phase, and a blazed grating phase. After blocking the undesired diffraction orders, the first diffracted order was carrying the imprinted information for further processing. A quarter waveplate was used to convert the linearly polarized beam to a circularly polarized beam. The phase profile loaded on the SLM was projected onto the sample plane through the two-lens imaging system comprising Lens 1(f = 500 mm) and an Objective lens (Olympus lucplanfln 20×, NA=0.45). An aperture-type near-field scanning optical microscope (NSOM) (NT-MDT NTEGRA Solaris) with an aluminum-coated fiber tip (tip aperture approximately 100 nm in diameter) was used to image the SPP fields on the top surface of the gold film. The distance between the sample and the fiber tip was controlled by a non-optical shear-force feedback mechanism, and was typically less than 20 nm. the resulting in-plane SPP field was collected by the probe and converted into guided modes within an optical fiber, the other end of which was connected to a photo-multiplier tube (PMT) for signal detection and amplification. The phase response of the liquid crystal molecules of the SLM varies with wavelengths, which was compensated for to produce the desired behavior of the whole-SPP fields by incorporating the reference phase response curve of the liquid crystal into the loaded phase pattern.

**S5: Additional results for dynamically-modulated SPP fields.**

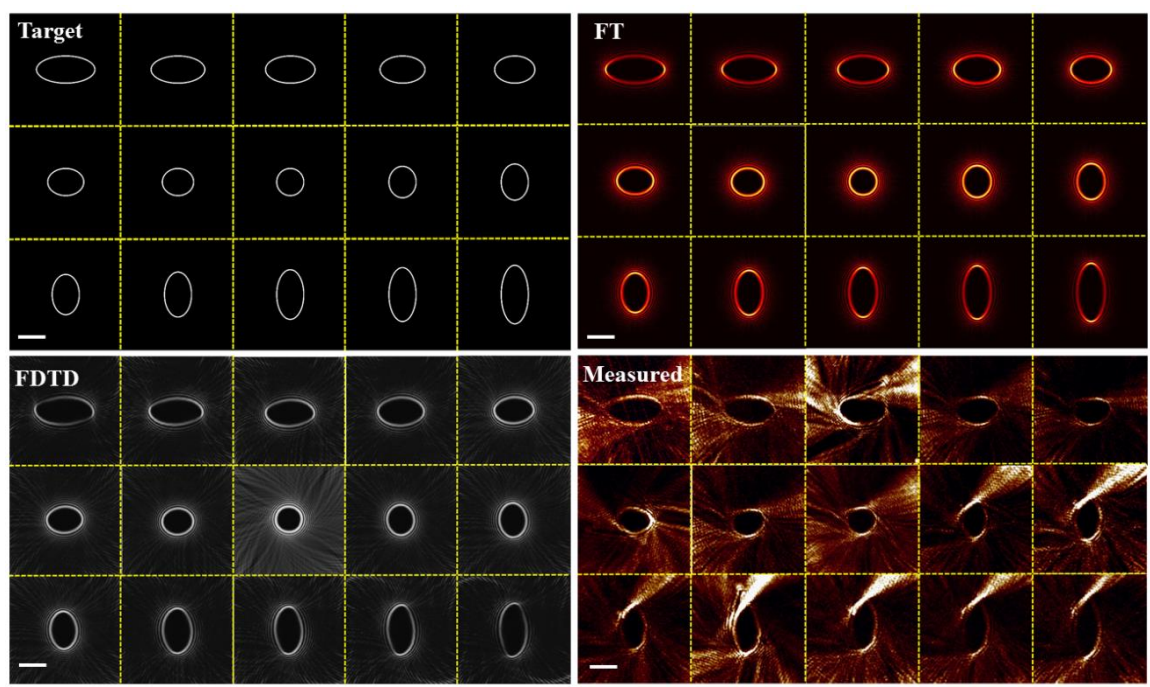

Figure S5 Results for dynamically-modulated SPP fields: Target section, FT section, FDTD section, and Measured section.

The target section shows the target patterns which we want to generate on the surface of the gold film. The eccentricities of the ellipses decrease from 0.9 to 0 and then increase from 0 to 0.9 with a change in the orientation of the major radius to the perpendicular axis. The FT section shows the fields calculated by performing a Fourier transform of the initial-launch-condition. The finite difference time domain (FDTD) calculating results are shown in the FDTD section. The final section is the 'Measured' section, which presents the NSOM-detected results of the in-plane components of the SPP fields on the gold film surface. It is a simple matter to modify the in-plane SPP fields by dynamically controlling the holograms loaded on the SLM.